\documentclass[lettersize,journal]{IEEEtran}
\usepackage{academicons}
\usepackage{array}
\usepackage[caption=false,font=normalsize,labelfont=sf,textfont=sf]{subfig}
\usepackage{textcomp}
\usepackage{stfloats}
\usepackage{url}
\usepackage{verbatim}
\usepackage{graphicx}
\usepackage{cite}
\usepackage[hidelinks]{hyperref}

\usepackage{multirow}%
\usepackage{amsmath,amssymb,amsfonts}%
\usepackage{amsthm}%
\usepackage{mathrsfs}%
\usepackage{xcolor}%
\usepackage{manyfoot}%
\usepackage{booktabs}%
\usepackage{algorithm}%
\usepackage{algorithmicx}%
\usepackage{algpseudocode}%
\usepackage{listings}%
\usepackage{lmodern}
\usepackage{pdflscape}

\begin{document}

\title{Towards Lensless Image Deblurring with Prior-Embedded Implicit Neural Representations in the Low-Data Regime}

\author{Abeer Banerjee and Sanjay Singh
}

\markboth{}%
{Shell \MakeLowercase{\textit{et al.}}: A Sample Article Using IEEEtran.cls for IEEE Journals}

\maketitle

\begin{abstract}
The field of computational imaging has witnessed a promising paradigm shift with the emergence of untrained neural networks, offering novel solutions to inverse computational imaging problems. While existing techniques have demonstrated impressive results, they often operate either in the high-data regime, leveraging Generative Adversarial Networks (GANs) as image priors, or through untrained iterative reconstruction in a data-agnostic manner. This paper delves into lensless image reconstruction, a subset of computational imaging that replaces traditional lenses with computation, enabling the development of ultra-thin and lightweight imaging systems. To the best of our knowledge, we are the first to leverage implicit neural representations for lensless image deblurring, achieving reconstructions without the requirement of prior training. We perform prior-embedded untrained iterative optimization to enhance reconstruction performance and speed up convergence, effectively bridging the gap between the no-data and high-data regimes. Through a thorough comparative analysis encompassing various untrained and low-shot methods, including under-parameterized non-convolutional methods and domain-restricted low-shot methods, we showcase the superior performance of our approach by a significant margin.
\end{abstract}

\begin{IEEEkeywords}
Lensless Imaging, Implicit Neural Representations, Computational Imaging,  Inverse Problems, Computational Photography 
\end{IEEEkeywords}

\section{Introduction}\label{sec1}
\IEEEPARstart{I}{n} the 21st century, computational imaging has undergone remarkable progress that was enabled by the advancement of computational capabilities, sensor technologies, and algorithmic innovations. This evolution has given rise to computational photography as a distinct discipline, harnessing computational methodologies to transcend the limitations inherent in traditional optical systems. Notably, recent years have witnessed a paradigm shift in the field, propelled by the emergence of generative adversarial networks (GANs) which have revolutionized inverse computational imaging tasks.

GANs have made a great impact in the field of inverse computational imaging by tackling complex tasks in low-level vision with remarkable efficacy. Tasks such as image denoising, super-resolution, deblurring, and inpainting have been significantly explored in the high-data regime, yielding impressive quantitative scores particularly when the problem is framed using a conditional GAN framework. However, these achievements are made possible by vast amounts of paired data and extensive training time.

Focusing specifically on lensless image reconstruction, our primary concern in this paper, the problem can be framed as an inverse computational imaging problem of image deblurring, where the blur kernel corresponds to the point spread function (PSF). Several approaches utilizing GANs for lensless image reconstruction have demonstrated noteworthy results, often producing outputs that are visually pleasing and occasionally even photorealistic. However, a significant limitation of these methods lies in their reliance on extensive training data and their PSF-agnostic nature.

One of the main drawbacks of existing GAN-based approaches for lensless image reconstruction is their inability to adapt to changes in the PSF. Even minor variations in the PSF can render these methods ineffective, thus limiting their practical utility in real-world scenarios. Moreover, the requirement for large amounts of training data poses practical challenges, particularly in domains where acquiring labeled datasets is labor-intensive or impractical.

Within the domain of lensless imaging, an intriguing approach involves the utilization of masks for modulating light in computational cameras. This approach can be dissected into two fundamental layers: the physical layer, which encompasses light-capturing components, and the digital layer, which encompasses algorithms for processing captured light to construct computational images. In the context of lensless imaging, our primary focus in this paper, the physical layer often relies on amplitude-modulating masks or phase-modulating masks.

\subsection{The Problem}
Lensless imaging systems, which eliminate the need for traditional lenses, can reconstruct images by replacing the lens with inverse computation. The primary challenge in lensless imaging is the recovery of high-fidelity images from measurements characterized by complex point spread functions (PSFs). This problem can be formulated as an image deblurring or deconvolution problem.

The forward imaging process in lensless imaging can be represented mathematically as:

\begin{equation}
y = A(x_0) + \eta
\label{eq1}
\end{equation}

In this equation, \(y\) represents the observed measurement or the captured image. The term \(A\) denotes the forward operator that describes the imaging system, which acts on the original scene or object \(x_0\) to produce the measurement. The term \(\eta\) represents the noise in the measurement, which can arise from various sources during the imaging process.

For lensless imaging, the forward operator \(A\) is typically described by the convolution operation with the PSF, denoted as \(k\):

\begin{equation}
A(x) = k \ast x
\label{eq2}
\end{equation}

Here, \(k\) is the blur kernel, or the PSF, of the imaging system, and \(\ast\) denotes the 2D convolution operation. The PSF characterizes how each point in the scene is blurred, effectively mapping the light distribution from the scene to the sensor. Consequently, the observed measurement \(y\) can be rewritten as:

\begin{equation}
y = k \ast x_0 + \eta
\label{eq3}
\end{equation}

The objective in lensless imaging is to solve the inverse problem, which involves reconstructing the original scene \(x_0\) from the measurement \(y\). \(A^{-1}\) would denote the inverse operation of the forward model. Directly solving the inverse problem is often ill-posed, especially in the presence of noise. To address this, regularization techniques are employed to stabilize the solution. The regularized inverse problem can be formulated as:

\begin{equation}
\hat{x} = \arg \min_x \left\{ \| y - k \ast x \|^2 + \lambda \mathcal{R}(x) \right\}
\label{eq5}
\end{equation}

In this formulation, \(\| y - k \ast x \|^2\) is the data fidelity term, ensuring that the reconstructed image \(x\) is consistent with the observed measurement \(y\). The term \(\mathcal{R}(x)\) represents the regularization term, which imposes prior knowledge or constraints on the solution \(x\). The parameter \(\lambda\) balances the data fidelity and regularization terms, allowing for control over the smoothness and accuracy of the reconstructed image.

Lensless imaging systems often employ sophisticated PSFs, such as random diffusers, to achieve high-resolution reconstructions. These systems use diffusers to scatter light in a controlled manner, creating a complex and random PSF. For instance, in the DiffuserCam system \cite{antipa2018diffusercam}, a random diffuser placed in front of the sensor produces a unique PSF for each point in the scene, enabling single-exposure 3D imaging. The complexity of the PSF in lensless systems presents both challenges and opportunities. While it can make the deblurring problem more difficult due to the intricate nature of the convolution operation, it also allows for enhanced imaging capabilities such as 3D reconstruction from a single 2D measurement.

Thus, the process of lensless imaging, viewed through the lens of image deblurring, underscores the importance of accurately modeling the PSF and effectively solving the resulting inverse problem to achieve high-quality image reconstruction. The advancements in computational methods and regularization techniques are pivotal in addressing these challenges and harnessing the full potential of lensless imaging systems.

\section{Related Works}
Significant research in the computational imaging community has focused on novel techniques for solving inverse problems, leveraging generative priors, untrained neural network priors, and unfolding networks. These advancements have significantly improved the capability to address inverse problems across various domains such as medical imaging, remote sensing, and computational photography. Our particular focus will be on untrained methods and methods that have attempted to solve inverse problems in the low-data regime.

\subsection{Generative Priors}
Generative priors, typically based on Generative Adversarial Networks (GANs) \cite{goodfellow2014generative}, are probabilistic models trained to generate images that reflect the data distribution of a large dataset. These models act as natural image priors for image restoration, and robust theories have been developed for solving inverse problems using these models, including compressed sensing \cite{bora2017compressed, DBLP:journals/corr/HandV17, Huang2018APC, hussein2020image, wu2019deep}, phase retrieval \cite{hand2018phase, Shamshad2018RobustCP}, and blind deconvolution \cite{Asim2018BlindID, Hand2019GlobalGF}. Despite their effectiveness, deep generative models have limitations such as the requirement for large amounts of training data and non-trivial representation error. PSF-agnostic generative models have been employed for lensless image reconstruction \cite{asif2016flatcam, antipa2018diffusercam}, though these models are camera-specific and require retraining for changes in PSF.

\subsection{Untrained Neural Network Priors}
Recent advancements have shown that randomly initialized neural networks can act as natural image priors without needing extensive training datasets, unlike traditional methods that require large data amounts. For instance, \cite{Ulyanov2017DeepIP} demonstrated that inverse problems like denoising, inpainting, and super-resolution can be solved by optimizing a convolutional neural network to fit a single image. \cite{heckel_deep_2018} introduced under-parameterized optimization for image compression using non-convolutional networks for linear inverse imaging. In lensless imaging, \cite{monakhova2021untrained} explored untrained networks for lensless imaging, \cite{banerjee2023physics} explored untrained reconstruction with both over-parameterized and under-parameterized neural networks, highlighting their potential. The dependency of deep-learning-based techniques on large labeled datasets poses challenges in fields such as medical imaging and microscopy, where untrained reconstruction can be improved in the low-data regime, i.e., using only 10 to 20 instances \cite{leong2019low}, or by using domain-restricted instances \cite{banerjee2023reconstructing}.

Most iterative algorithms are inherently data-agnostic in solving inverse problems. Classical Maximum a Posteriori (MAP) reconstruction algorithms, such as those based on the Alternating Direction Method of Multipliers (ADMM) and Total Variation (TV) regularization, have been extensively researched for lensless image reconstruction \cite{Rudin1992NonlinearTV, Beck2009AFI, boyd2011distributed}. However, there exists a trade-off between reconstruction quality and computational time. Well-regularized untrained neural networks have successfully addressed various inverse problems without training data \cite{Heckel2019RegularizingLI, jagatap2019algorithmic, Veen2018CompressedSW}. Bridging the gap between no-data and high-data regimes remains challenging. \cite{leong2019low} proposed a low-shot learning technique that uses an untrained neural network to leverage benefits from both regimes, applicable to tasks like colorization and compressed sensing.

The recent advancements in physics-driven learning-based algorithms have shown impressive performance and faster convergence in various inverse problems \cite{wen2023physics, poirot2019physics, deng2020interplay, karniadakis2021physics}. These approaches incorporate physical models directly into the learning process, enhancing the interpretability and efficiency of the reconstructions.

In lensless imaging, traditional methods struggle with the complexity of the point spread function (PSF), which varies between setups. While generative approaches using GANs can offer high-quality reconstructions, they require extensive training data and are often PSF-agnostic, limiting their practicality when the PSF changes. Our work bridges this gap by leveraging implicit neural representation learning for lensless image reconstruction, demonstrating improved performance in the low-data regime, and enhancing untrained models with minimal domain-specific examples. Our contributions in this paper can be highlighted as follows:

\begin{itemize}
\item We are the first to leverage Implicit Neural Representations (INRs) for lensless image reconstruction, introducing a novel approach to address the challenges posed by complex Point Spread Functions (PSFs) and limited data availability.

\item We utilize the concept of prior embedding to enhance the convergence speed and reconstruction quality of untrained INRs. This technique significantly improves reconstruction performance, particularly in scenarios with minimal domain-specific data.

\item We provide a comprehensive evaluation of our method against established techniques, demonstrating superior performance in terms of both PSNR, SSIM, and qualitative visual quality. We propose the Under-Parameterization Ratio (UPR) which provides insights into the degree of under-parameterization in the network, offering valuable guidance for model comparison in the untrained and low-data regimes.
\end{itemize}

\section{Methodology}
In this section, we present the methodology employed for lensless image reconstruction using Implicit Neural Representations (INRs). The section begins with the context and motivation for the chosen techniques. Following this, we introduce the concept of under-parameterizations, and we mathematically formulate the forward imaging model, which describes the imaging process and forms the basis for the inverse problem of image reconstruction. We incorporated a fast and accurate version of the forward model into the untrained optimization loop which includes the point spread function (PSF) and the intermediate reconstruction generated by the network. Next, we introduce the network architecture used for the implicit neural representation of the lensed images, highlighting its ability to represent continuous image signals and the benefits it brings to the reconstruction task. 

The section on untrained optimization explains the strategies used to optimize the network parameters without prior training on large datasets. This includes the objective functions, regularization techniques, and optimization algorithms employed to achieve effective reconstructions. We also discuss how low-shot learning is incorporated to improve performance and convergence in the low-data regime.

\subsection{Under-Parameterized Networks} An under-parameterized network has significantly fewer trainable parameters than the amount of available data or the intrinsic dimensionality of the data manifold. This contrasts with over-parameterized networks, which have an excess of parameters and often require extensive training data for generalization.

An under-parameterized network can be described by its architecture. Considering a network \( \mathcal{M}_{\theta} \) with parameters \( \theta \), taking a fixed, random input \( z \) (latent code) to generate an output. Mathematically, the optimization process for fitting an image \( x \) of size \( m \times n \) can be expressed as:

\begin{equation}
\theta^* = \arg\min_{\theta} \mathcal{L}(\mathcal{M}_{\theta}(z), x)
\end{equation}

where \( \mathcal{L} \) is the loss function measuring the discrepancy between the network output and the target image \( x \).

For a linear under-parameterized network, the relationship between the target image \( x \in \mathbb{R}^d \) and the network can be expressed as:

\begin{equation}
x \approx Wz
\end{equation}

where \( W \in \mathbb{R}^{d \times k} \) with \( k \ll d \), and \( z \in \mathbb{R}^k \) is the low-dimensional latent code. The optimization problem then becomes:

\begin{equation}
\min_{W, z} \| x - Wz \|_2^2
\end{equation}

Under-parameterized networks are data-efficient, requiring less data for training and preventing overfitting due to their fewer parameters. This makes them ideal for scenarios with limited data availability, providing interpretable learned representations useful in applications like medical imaging and scientific research. The Deep Decoder is a notable example of an under-parameterized, non-convolutional network designed for concise image representations \cite{heckel_deep_2018}. It uses a simple architecture with fewer parameters, achieving competitive results in image reconstruction tasks without extensive training data. The implicit regularization from its under-parameterized nature effectively prevents overfitting, promoting high-quality image recovery from limited data.

\section{Implicit Neural Representations}
Implicit Neural Representations (INRs) have gained significant attention in the field of image reconstruction, particularly for solving inverse problems such as image deblurring. Unlike traditional explicit representations that rely on discrete grid-based data structures, INRs model continuous signals as functions parameterized by neural networks. This paradigm shift allows for more flexible and efficient representations of high-dimensional data, offering several advantages in terms of memory efficiency, smoothness, and the ability to handle sparse or irregularly sampled data. Implicit Neural Representations are continuous functions parameterized by neural networks that map spatial coordinates to signal values. For an image \( x \) defined over a domain \( \Omega \subset \mathbb{R}^2 \), an INR \( \mathcal{M} \) can be formulated as:

\begin{equation}
\mathcal{M}_\theta: \mathbb{R}^2 \rightarrow \mathbb{R}^3, \quad (u, v) \mapsto \mathcal{M}_\theta(u, v)   
\end{equation}

where \( \theta \) denotes the parameters of the neural network, \( \mathbb{R}^2\)) corresponds to the pixel coordinate space and \( \mathbb{R}^3 \) corresponds to the RGB value of the pixel. The forward problem can be described by the equation \( y = A(x) + \eta \) as discussed in detail in the Sec. \ref{sec_fm}, where \( y \) is the observed blurred image, \( A \) is the forward operator modeling the blur (often a convolution with a point spread function \( k \)), and \( \eta \) represents noise. Using an INR, we aim to reconstruct the deblurred image \( x \) by learning a neural network \( \mathcal{M}_\theta \) such that:

\begin{equation}
x(u, v) \approx \mathcal{M}_\theta(u, v)
\end{equation}

The deblurring problem can then be reformulated as an optimization problem:

\begin{equation}
\theta^* = \arg\min_\theta \lvert \lvert A\mathcal{M}_\theta - y \rvert \rvert_{2}^{2} + \lambda \mathcal{R}(\theta)
\end{equation}

where \( \mathcal{L} \) is a loss function measuring the discrepancy between the blurred observation \( y \) and the predicted blur from \( \mathcal{M}_\theta \), \( \mathcal{R} \) is a regularization term, \( \lambda \) is a regularization parameter.

The forward operator \( A \) for the deblurring problem can typically be approximated using a convolution operation under certain limits discussed in Sec. \ref{sec_fm}:

\begin{equation}
A\mathcal{M}_\theta(u, v) = k \ast \mathcal{M}_\theta(u, v)    
\end{equation}

where \( k \) is the blur kernel (PSF), and \( \ast \) denotes the convolution operation. The same mathematical framework can be extended for the cases of other inverse imaging problems where INRs can be used to represent images as continuous functions, significantly reducing memory usage compared to traditional grid-based representations, which is particularly beneficial for high-resolution images or 3D volumes. By leveraging neural networks, INRs inherently produce smooth and continuous signal representations for high-quality reconstructions. The ability to naturally handle sparse and irregularly sampled data makes INRs suitable for applications where data acquisition is challenging or expensive. The parameterization of signals using neural networks allows for compact representations, useful for storage and transmission. However, training INRs can be computationally intensive, particularly for high-dimensional data, requiring significant processing power and time. INRs can also be sensitive to the choice of network architecture and hyperparameters, thus we have performed extensive hyperparameter tuning and ablation study to finalize the network architecture. The performance of INRs can be highly dependent on the initialization of network parameters, impacting convergence and final reconstruction quality. To tackle this issue, we use prior embedding as a low-shot mechanism to perform domain-restricted reconstruction of lensless images. This significantly reduces the convergence time and improves the quality of the reconstructed image.

\subsection{The Forward Model}\label{sec_fm}
The general equation for the forward model \( \mathbb{A} \) is given by:

\begin{equation}
y = \mathbb{A} \cdot x
\end{equation}

Modeling the PSF as shift-invariant, where the patterns generated by off-axis point sources are approximated as laterally shifted versions of the on-axis PSF, significantly reduces computational complexity, storage requirements, and calibration efforts. However, this approach introduces non-idealities because it relies on physically inexact PSFs as argued by \cite{zeng2021robust}. The on-axis PSF, obtained either through experimental calibration or simulation based on the mask, is inherently error-prone. Consequently, the shift-invariant assumption does not perfectly represent the true imaging process and can lead to inaccuracies in the reconstructed images. The forward model refers to the forward problem of the lensless image formation process. In the specific inverse problem of lensless image deblurring, the lensless image formation process can be approximated using a convolution operation. Here, the lensed image \( x \) is convolved with the point spread function (PSF) \( k \) to obtain the lensless image \( y \). This relationship is expressed as:

\begin{equation}
y = \boldsymbol{k} \ast x + \eta
\end{equation}

In this equation, \( x \) is the input image (lensed image), \( k \) represents the blurring effect caused by the absence of a lens, and \( \eta \) represents noise. The convolution operation effectively models the spreading and overlap of light from each point in the scene as it passes through the random diffuser. 

During the training process, the network prediction is in the lensed domain and is referred to as the intermediate lensed image. To simulate the forward process, we computationally obtain the intermediate lensless image by convolving the intermediate lensed image with the known PSF and adding a small amount of additive Gaussian noise. This approximates the forward process accurately and efficiently. Given that the forward process is part of the training loop, this approximation must be both fast and accurate. Hence, we resort to FFT convolution for its computational efficiency. Using FFT convolution, the lensless image can be expressed as:

\begin{equation}
y = \mathcal{F}^{-1}(\mathcal{F}(x) \cdot \mathcal{F}(k))
\end{equation}

where \( \mathcal{F} \) denotes the Fourier transform and \( \mathcal{F}^{-1} \) denotes the inverse Fourier transform. The symbol \( \cdot \) represents element-wise multiplication in the Fourier domain.

To enhance the robustness of the reconstruction process, we add a small amount of Gaussian noise to the computed lensless image. The speed of the forward model is of paramount importance to the training pipeline since it is used for calculating the cycle consistency loss within the training loop. This ensures that the model can efficiently and accurately approximate the lensless imaging process during training.

\subsection{Reconstruction with INRs}
The INR \( \mathcal{M}_{\theta} \) aims to reconstruct the original image \( x \) from \( y \) by learning the mapping:

\begin{equation}
x \approx \mathcal{M}_{\theta}(u, v)
\end{equation}

Here, \( (u, v) \) are the continuous spatial coordinates of the image domain. \cite{sitzmann2020implicit} suggested the use of periodic activation functions, such as sinusoidal activations, which enable the network to represent high-frequency details more effectively. Our implementation of INRs for image deblurring is based on the SIREN architecture proposed by \cite{sitzmann2020implicit}, which utilizes sinusoidal activation functions to effectively capture high-frequency details. The SIREN architecture is defined by a sequence of layers, each comprising a linear transformation followed by a sinusoidal activation function. The sinusoidal activation function can be expressed as:

\begin{equation}
\sigma(z) = \sin(\omega_0 z)    
\end{equation}

where \(\omega_0\) is a frequency scaling parameter. Each layer in the SIREN network performs a linear transformation followed by sinusoidal activation. Each layer in the SIREN network performs Linear Transformation: The input is multiplied by a weight matrix and added with a bias vector, and Sinusoidal Activation: The result of the linear transformation is passed through a sinusoidal activation function. The output of the \( i \)-th layer, denoted as \( \mathbf{h}^{(i)} \), can be represented mathematically as:
\begin{equation}
\mathbf{h}^{(i)} = \sin(\omega_0^{(i)} (\mathbf{W}^{(i)} \mathbf{h}^{(i-1)} + \mathbf{b}^{(i)}))    
\end{equation}
where \( \omega_0^{(i)} \) is the frequency scaling parameter for the \( i \)-th layer. \( \mathbf{W}^{(i)} \) is the weight matrix of the \( i \)-th layer. \( \mathbf{b}^{(i)} \) is the bias vector of the \( i \)-th layer. \( \mathbf{h}^{(i-1)} \) is the output of the previous layer.

The weights and biases of the SIREN network are initialized differently based on whether the layer is the first layer or subsequent layers. For the first layer, the weights are initialized uniformly in the range \( \left[ -\frac{1}{d_{\text{in}}}, \frac{1}{d_{\text{in}}} \right] \), where \( d_{\text{in}} \) is the number of input features. For subsequent layers, the weights are initialized uniformly in the range \( \left[ -\frac{\sqrt{6}}{\omega_0^{(i)} \sqrt{d_{\text{in}}}}, \frac{\sqrt{6}}{\omega_0^{(i)} \sqrt{d_{\text{in}}}} \right] \).

The SIREN network consists of an input layer, several hidden layers, and an output layer. The input layer receives the input coordinates, and the output layer produces the predicted pixel values. Each hidden layer applies the sinusoidal activation function to the output of the previous layer. The input coordinates represent the pixel coordinates of the blurred image, and the output represents the pixel values of the deblurred image. By learning the continuous function parameterized by the SIREN network, the INR effectively models the underlying clean image from the blurred observation subject to the physics-informed forward loss function.

\begin{figure}[h]
\centering
\includegraphics[width=0.49\textwidth]{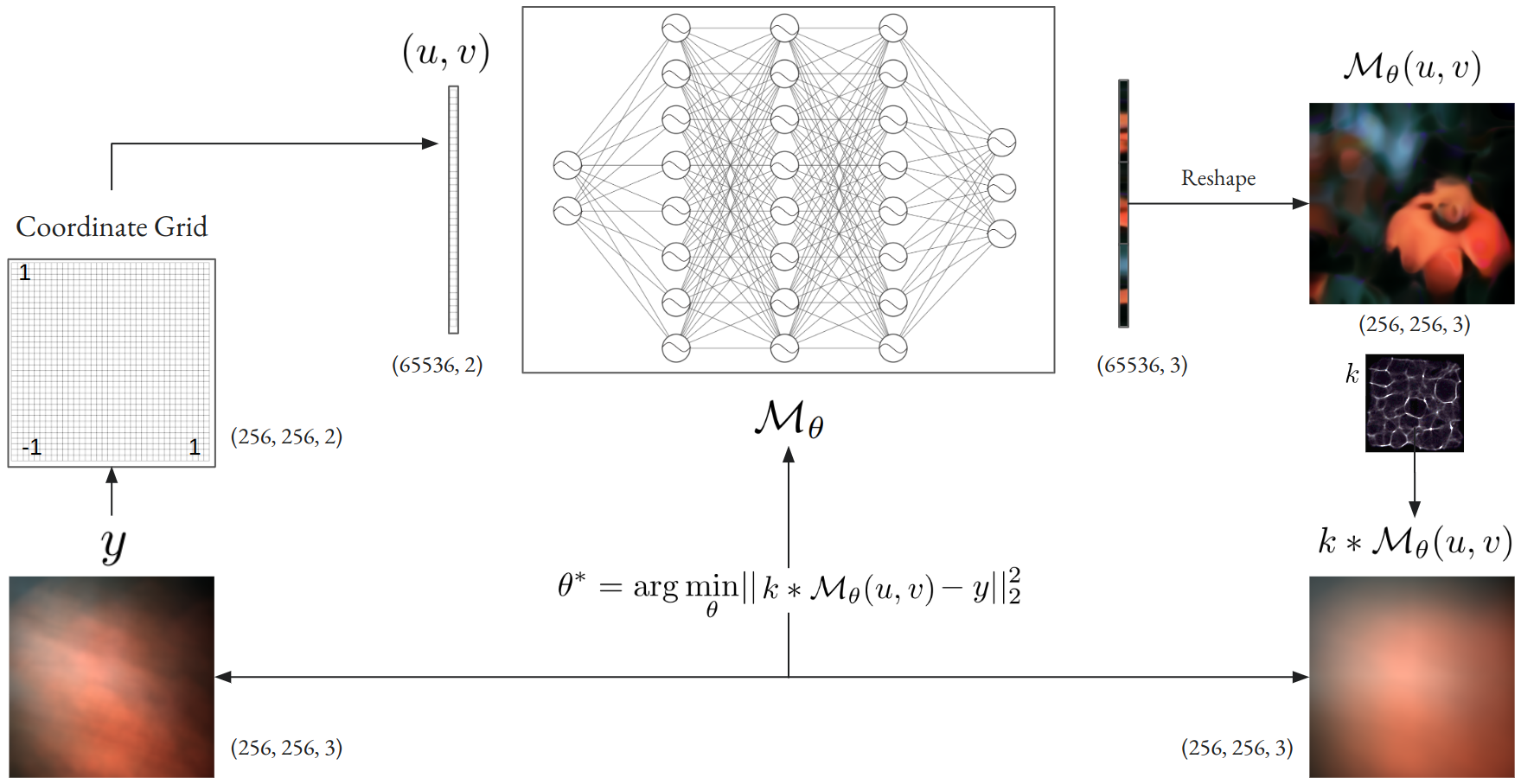}
\caption{The complete untrained iterative optimization procedure using the implicit neural representation framework has been illustrated here. We coordinate input \((u, v)\) for the MLP \(\mathcal{M}_\theta\). Intermediate reconstruction \(\mathcal{M}_\theta(u, v)\), which is then convolved with the known point spread function (PSF) \(k\) to produce the intermediate lensless image \(k \ast \mathcal{M}_\theta(u, v)\). $L_{MSE}$ between this intermediate lensless image and the original lensless image \(y\) is backpropagated to update \( \mathcal{M}_\theta\).}
\label{trainfig}
\end{figure}

\subsection{Untrained Iterative Optimization}
We propose an untrained iterative optimization approach for Implicit Neural Representations (INRs) to address the task of image deblurring. The algorithm, outlined in Algorithm \ref{alg1}, iteratively refines the predicted deblurred image using the forward model and updates the network parameters based on the computed loss.
 
\begin{algorithm}
\caption{Untrained Iterative Optimization for INRs}
\label{alg1}
\begin{algorithmic}[1]
\State \textbf{Input:} Blurred image $y$, Initial random coordinates $u, v$, Total optimization steps $T$, Steps until summary $S$, Learning rate $\eta$, Loss criterion $\mathcal{L}$
\State \textbf{Output:} Deblurred image $\tilde{x}$
\State Initialize Network Parameters: $\mathcal{M}_{\theta}$
\For{$t \gets 1$ \textbf{to} $T$}
    \State Compute predicted deblurred image: $\tilde{x}(u, v) = \mathcal{M}_{\theta}(u, v)$
    \State Compute forward model prediction: $\tilde{y} = \boldsymbol{k} \ast \tilde{x}(u, v)$
    \State Compute loss: $loss = \mathcal{L}(\tilde{y}, y)$
    \State Update $\theta$ using gradient descent: $\theta \gets \theta - \eta \cdot \nabla_{\theta} \mathcal{L}$
\EndFor
\end{algorithmic}
\end{algorithm}

In this algorithm, $\mathcal{M}_{\theta}$ represents the INR network with parameters $\theta$, which is initialized randomly. The input to the algorithm includes the blurred image $y$, initial random coordinates $u, v$, total optimization steps $T$, steps until summary $S$, learning rate $\eta$, and the loss criterion $\mathcal{L}$. The output is the deblurred image $\tilde{x}$. At each iteration $t$, the algorithm computes the predicted deblurred image $\tilde{x}(u, v)$ using the INR network. Then, it computes the forward model prediction $\tilde{y}$ by convolving the predicted deblurred image with the point spread function (PSF) $\boldsymbol{k}$. The loss is calculated between the predicted and ground truth blurred images using the $L_1$ loss function. Finally, the network parameters $\theta$ are updated using gradient descent to minimize the loss. 

As illustrated in Fig. \ref{trainfig}, we begin by creating and flattening a coordinate grid scaled between -1 and +1, which serves as the coordinate input \((u, v)\) for the MLP \(\mathcal{M}_\theta\). The output of the MLP is reshaped to form the image \(\mathcal{M}_\theta(u, v)\), which is then convolved with the known point spread function (PSF) \(k\) to produce the intermediate lensless image \(k \ast \mathcal{M}_\theta(u, v)\). The mean-squared error between this intermediate lensless image and the original lensless image \(y\) is calculated and backpropagated to update the parameters \(\theta\) of the MLP. This process, known as physics-informed forward loss optimization, incorporates the known PSF into the optimization loop to guide the reconstruction. 
\begin{figure*}
\centering
\includegraphics[width=\textwidth]{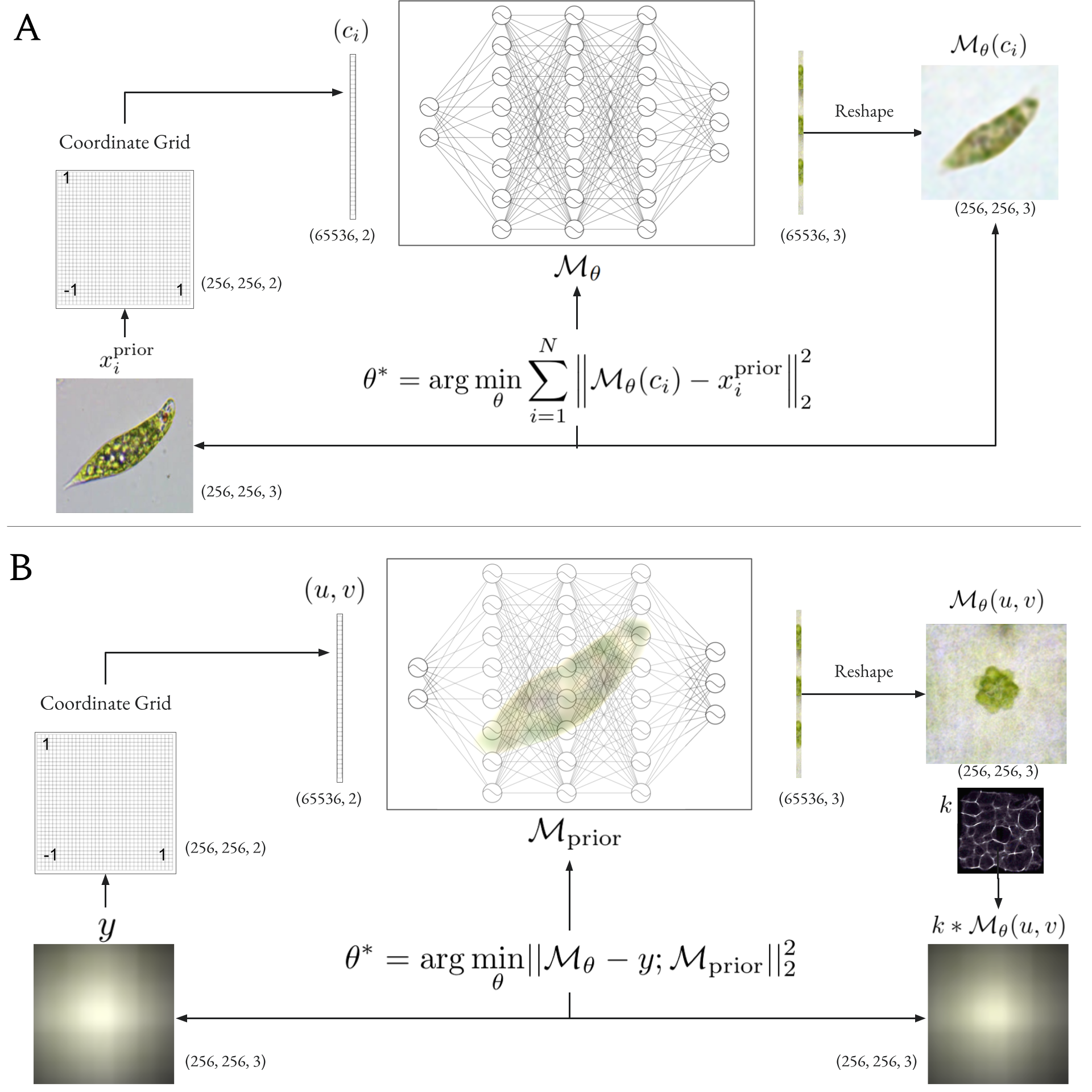}
\caption{\textbf{A.} Prior-Embedding process: A domain-restricted image is represented using the INR simply by optimizing the MSE between the network output and the image.
\textbf{B.} Prior-Embedded Physics-informed Inverse Imaging: The prior embedded INR $\mathcal{M}_{prior}$ is the starting point of the physics-informed optimization process that requires the PSF.}
\label{priorem}
\end{figure*}
\subsection{Prior Embedded Reconstruction} To cope with the problem of bad initialization of the INR which leads to a longer convergence time, we use the concept of prior embedding inspired by \cite{shen2022nerp}. This process utilizes a coordinate-based multi-layer perceptron (MLP), denoted as \(\mathcal{M}_{\theta}\), to map spatial coordinates to corresponding intensity values in the prior image \(x_{\text{prior}}\). The MLP performs the mapping \(\mathcal{M}_{\theta}: c_i \rightarrow x_i^{\text{prior}}\), where \(c_i\) represents the spatial coordinates and \(x_i^{\text{prior}}\) represents the intensity values at those coordinates. The prior image provides a set of coordinate-intensity pairs \(\{(c_i, x_i^{\text{prior}})\}_{i=1}^{N}\), with \(N\) being the total number of pixels in the image. The MLP is initialized randomly and optimized to minimize the mean squared error between its output and the intensity values of the prior image. This optimization is mathematically represented by the objective function:
\begin{equation}\label{eq13}
\theta^* = \arg\min_{\theta} \sum_{i=1}^{N} \left\| \mathcal{M}_{\theta}(c_i) - x_i^{\text{prior}} \right\|_2^2.    
\end{equation}

After the optimization process, the MLP network \(\mathcal{M}_{\theta^*}\) encodes the internal information of the prior image. For clarity, this optimized MLP with embedded prior information is referred to as \(\mathcal{M}_{\text{prior}}\), such that \(x_{\text{prior}} = \mathcal{M}_{\theta^*} = \mathcal{M}_{\text{prior}}\).

We first embed the prior image as the network weights, thereby initializing the network using Eq. \ref{eq13}. The optimization objective, considering the forward model \(k \ast\) and the prior-embedded network \(\mathcal{M}_{\text{prior}}\), is formulated as follows:

\[
\theta^* = \arg\min_{\theta} \lvert \lvert \mathcal{M}_{\theta} - y; \mathcal{M}_{\text{prior}} \rvert \rvert_2^2,
\]
where \(x^* = \mathcal{M}_{\theta^*}\). The network \(\mathcal{M}_{\theta}\) is trained by minimizing the L2-norm loss, starting from the initialization provided by the prior-embedded network \(\mathcal{M}_{\text{prior}}\). The complete process of prior-embedded untrained optimization has been illustrated as a two-step process in Fig. \ref{priorem}.

\section{Results}
We compare the reconstruction performance of untrained INRs against the Deep Decoder architecture. The Deep Decoder serves as an appropriate benchmark due to its established efficacy in solving various inverse problems, particularly in image reconstruction tasks. Deep Decoder utilizes a decoder-based approach without the need for pre-training on large datasets, making it a relevant point of comparison for our untrained INR method. Both methods leverage the inherent structure of neural networks to perform image deblurring, providing a fair basis for evaluating their relative strengths in terms of reconstruction quality and computational efficiency. We also perform a detailed visual comparison of our proposed approach against the untrained approach of Deep Decoder and other established methods previously employed for lensless imaging. All results were generated using an NVIDIA RTX 3090 graphics processor.

\subsection{Quantitative Results}
\begin{figure*}[t]
\centering
\includegraphics[width=\textwidth]{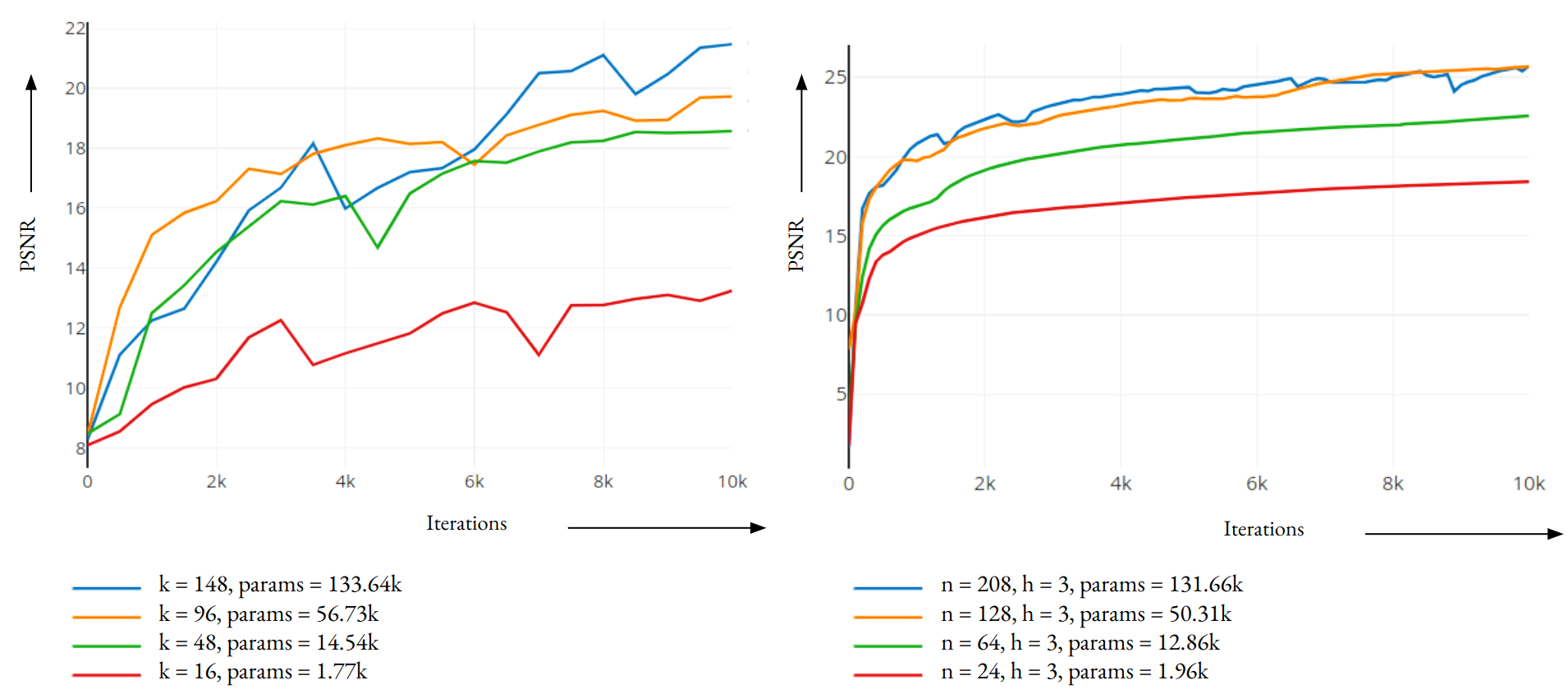}
\caption{PSNR versus iteration plots. The left plot depicts the results from the Deep Decoder framework, while the right plot shows the outcomes of our method. The parameter details have been highlighted below each plot.}
\label{qres}
\end{figure*}

To quantitatively assess the performance, we employ metrics such as Peak Signal-to-Noise Ratio (PSNR) and Structural Similarity Index Measure (SSIM). PSNR is widely used to measure the quality of reconstruction by comparing the maximum possible power of a signal to the power of corrupting noise. SSIM, on the other hand, evaluates the perceptual similarity between images, taking into account changes in structural information, luminance, and contrast. 

We utilize the testing images from the DiffuserCam dataset to perform a quantitative comparison. The images were resized to $256\times256$, so the resulting parameter count for image representation is $256\times256\times3=196,608$, i.e., around 196.6k. So, we chose hyperparameter combinations which resulted in parameter counts of the network being less than 196k making the optimization under-parameterized. Our results plotted in Fig. \ref{qres} demonstrate that the untrained INR achieves superior PSNR and SSIM values compared to the Deep Decoder. The untrained INR method converges faster and with fewer parameters, indicating effectiveness. The performance improvement can be attributed to the continuous function representation and periodic activation functions used in the INR, which provide a more compact and expressive representation of the image. 

Networks with too few parameters struggled to preserve structural information. However, as the parameter count was gradually increased, the reconstructed image began to retain both structure and color, resulting in a high-fidelity image. To analyze the impact of under-parameterization on the reconstructed output, we introduce a new metric called the Under-Parameterization Ratio (UPR). UPR is defined as the ratio between the dimensionality of the RGB image space and the number of parameters in the network being optimized. So, essentially, a UPR strictly greater than 1 would imply under-parameterization. For the analysis, we fix the hidden layer count of the MLP at 3 and vary the number of nodes in the case of Untrained INR, and for the modified Deep Decoder, we vary $k$ while keeping the layer count fixed at 5. We plot the variation of the SSIM against UPR for fixed numbers of iterations and see an obvious trend of increasing SSIM with decreasing UPR in both Deep Decoder and our untrained INR and the results have been presented in Table \ref{tab0}.


\begin{table*}[h]
\centering
\caption{This table shows the variation of SSIM with Under-Parameterization Ratio (UPR) for fixed numbers of iterations. MDD refers to the Modified Deep Decoder and INR is the proposed untrained INR method.}
\label{tab0}
\resizebox{\textwidth}{!}{%
\begin{tabular}{cc|cc|cc|cc|cc}
\hline
\hline
\multicolumn{2}{c|}{UPR} & \multicolumn{2}{c|}{Params} & \multicolumn{2}{c|}{2k SSIM} & \multicolumn{2}{c|}{6k SSIM} & \multicolumn{2}{c}{10k SSIM}\\
\hline
MDD & INR & MDD & INR & MDD & INR & MDD & INR & MDD & INR\\
\hline
1.47 (k=148)&1.49 (n=208)    & 133.64k&131.66k    &0.53&0.76  &0.69&0.83	&0.76&0.86\\
3.46 (k=96)&3.90 (n=128)     & 56.73k&50.37k      &0.59&0.72	&0.68&0.78	&0.73&0.85\\
13.52 (k=48)&15.28 (n=64)    & 14.54k&12.86k      &0.50&0.55	&0.60&0.70	&0.70&0.75\\
29.67 (k=32)&26.78 (n=48)    & 6.62k&7.34k        &0.47&0.51	&0.52&0.61	&0.66&0.69\\
51.46 (k=24)&58.51 (n=32)    & 3.82k&3.36k        &0.41&0.48	&0.44&0.54	&0.50&0.58\\
111.07 (k=16)&100.30 (n=24)  & 1.78k&1.94k        &0.38&0.37	&0.43&0.46	&0.47&0.50\\   
\hline
\hline
\end{tabular}%
}
\end{table*}


In the case of prior-embedded reconstruction, we observe significantly faster convergence and improved scores in both evaluation metrics. For a fair comparison, we benchmark our prior-embedded INR against the method presented in \cite{banerjee2023reconstructing}. In their work, the authors performed domain-restricted low-shot reconstruction of lensless images by providing 5 to 10 example images from the same domain to a decoder network. First, they performed low-shot training to obtain jointly optimized latent code and decoder weights. They then initialized the decoder with the optimized weights and used the mean-average latent code as its input, then the inverse problem of lensless image deblurring was performed via untrained iterative optimization.    
\begin{figure*}
\centering
\includegraphics[width=\textwidth]{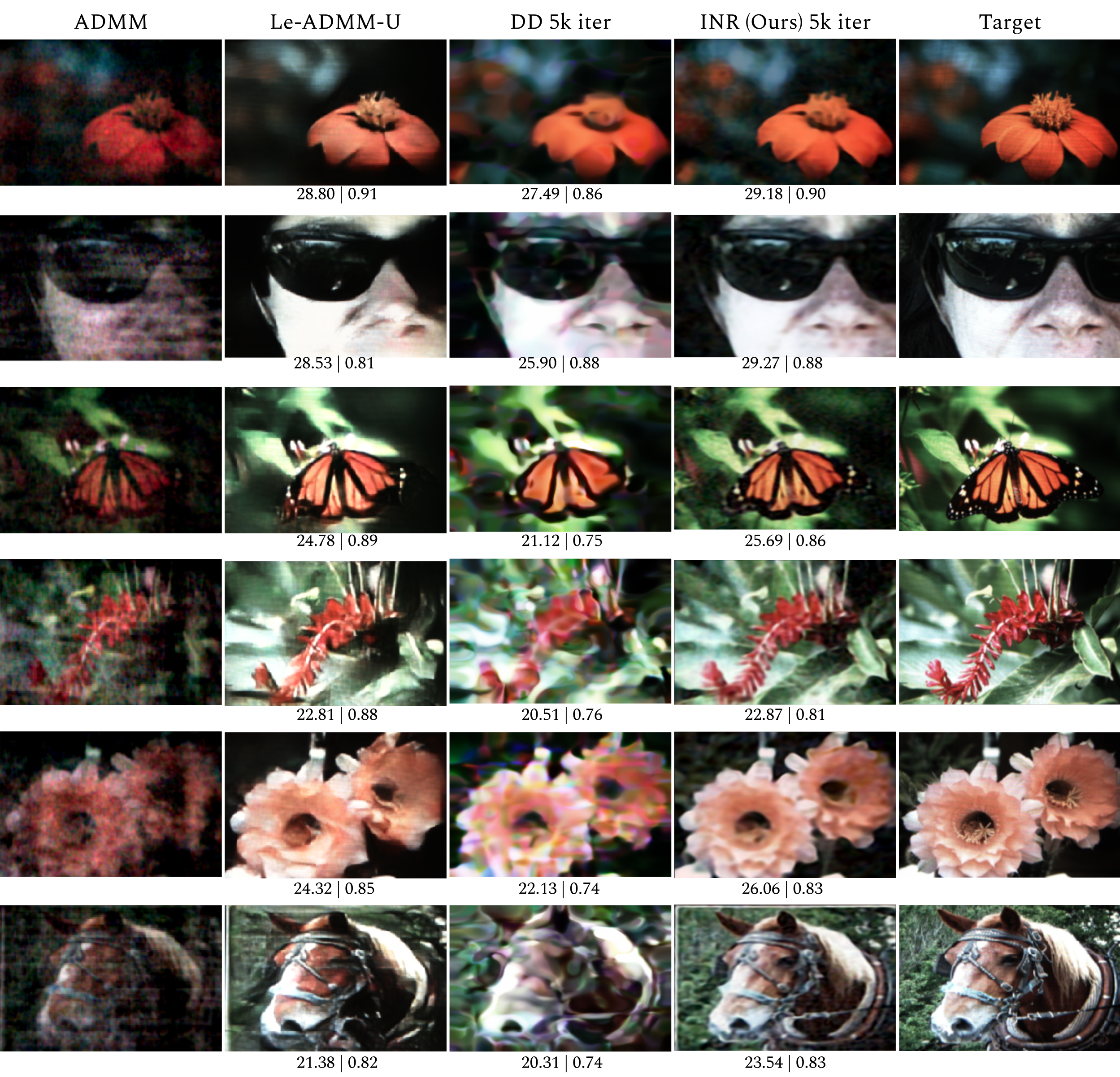}
\caption{Visual comparison of reconstruction results of existing methods like the ADMM\cite{antipa2018diffusercam} in the extreme left column, followed by Le-ADMM-UNet \cite{monakhova2019learned}, Deep Decoder \cite{heckel_deep_2018} modified for lensless image deblurring by \cite{banerjee2023physics}, against our untrained INR. The quantitative scores, presented in PSNR (in dB) | SSIM format, are provided below each image.}
\label{viscomp}
\end{figure*}
\subsection{Visual Comparison}
To support the quantitative evaluation of our method, we perform an extensive visual comparison in this section. We compare the reconstruction performance of our method against established methods such as ADMM \cite{boyd2011distributed} and Learned-ADMM-UNet \cite{monakhova2019learned}. Also, we conduct a visual comparison with the Deep Decoder \cite{heckel_deep_2018}, modified for lensless imaging by \cite{banerjee2023physics}. For a fair comparison, we have chosen to compare only untrained methods for lensless imaging, with the exception of Learned-ADMM-UNet, which is data-driven but included for its relevance.

The MLP used for the implicit neural representation had three hidden layers with 208 nodes, totaling 131.66k parameters and an under-parameterization ratio (UPR) of 1.49. We fixed the iteration count at 5000 for both the untrained INR and the modified Deep Decoder. The latter used k=148, resulting in a 133.64k parameter network with a UPR of 1.47. The ADMM method was run for 100 iterations, but it performed poorly compared to the other methods.

As illustrated in Fig. \ref{viscomp}, our Untrained INR consistently outperformed all the compared methods in nearly all the images. However, a closer visual inspection reveals that the Learned-ADMM-Unet slightly excelled in structure restoration. Clear improvements in reconstruction performance were observed against the modified Deep Decoder and notably against the ADMM. This demonstrates the robustness and efficacy of our approach in achieving high-quality reconstructions in lensless imaging.

\begin{figure*}[t]
\centering
\includegraphics[width=\textwidth]{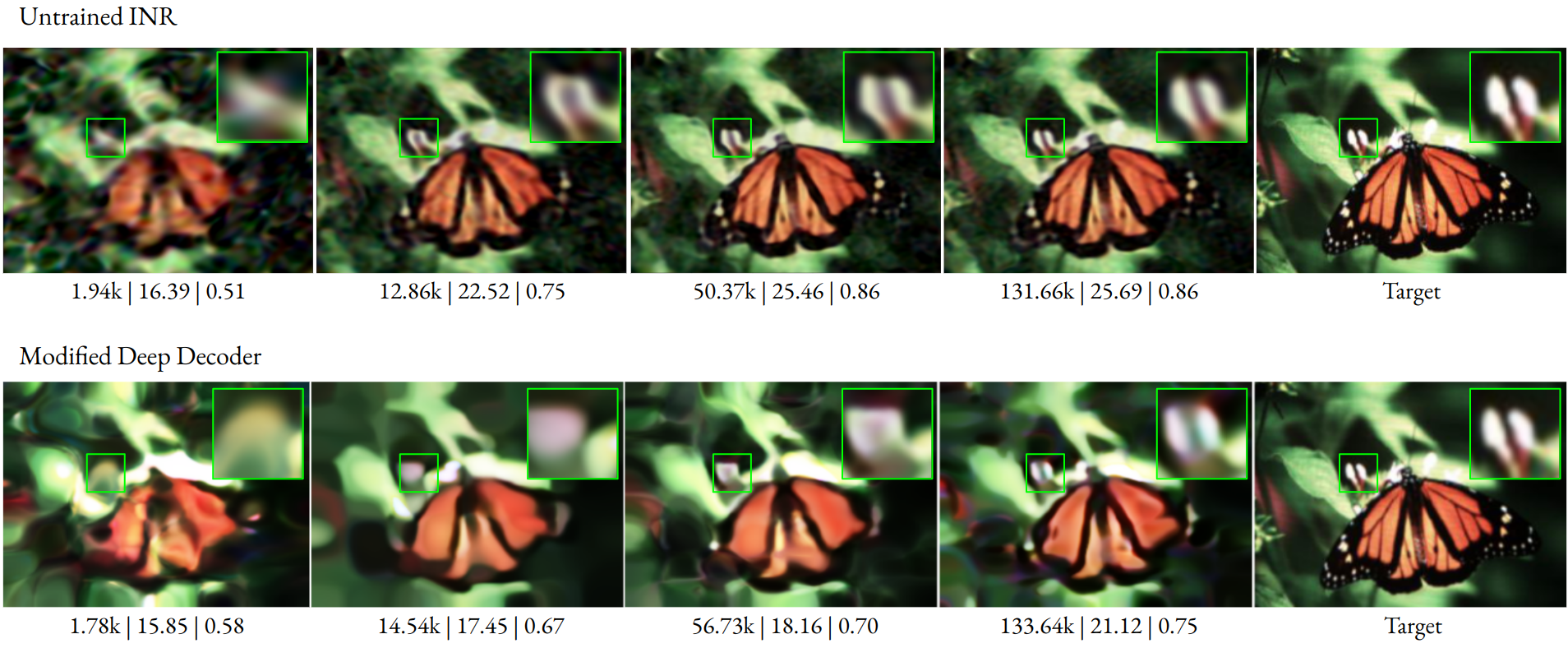}
\caption{Visual comparison of reconstruction results of our untrained INR against a modified Deep Decoder. The parameter count of the network and the quantitative scores, presented in Parameter Count | PSNR (in dB) | SSIM format, are provided below each image.}
\label{vresun}
\end{figure*}

We conducted an ablation study to evaluate the performance of untrained INRs with MLPs of varying under-parameterization ratios (UPRs). We compared these results against those obtained using the modified Deep Decoder with networks having similar UPR values. The reconstructed outputs from this study are illustrated in Fig. \ref{vresun}. It is evident from the figure that, for a given UPR, the untrained INR consistently outperforms the modified Deep Decoder both visually and quantitatively. The superior performance of the untrained INR is highlighted by clearer structural details and higher quantitative metrics, demonstrating the effectiveness of our approach in producing high-quality reconstructions.

\begin{figure}[t]
\centering
\includegraphics[width=0.5\textwidth]{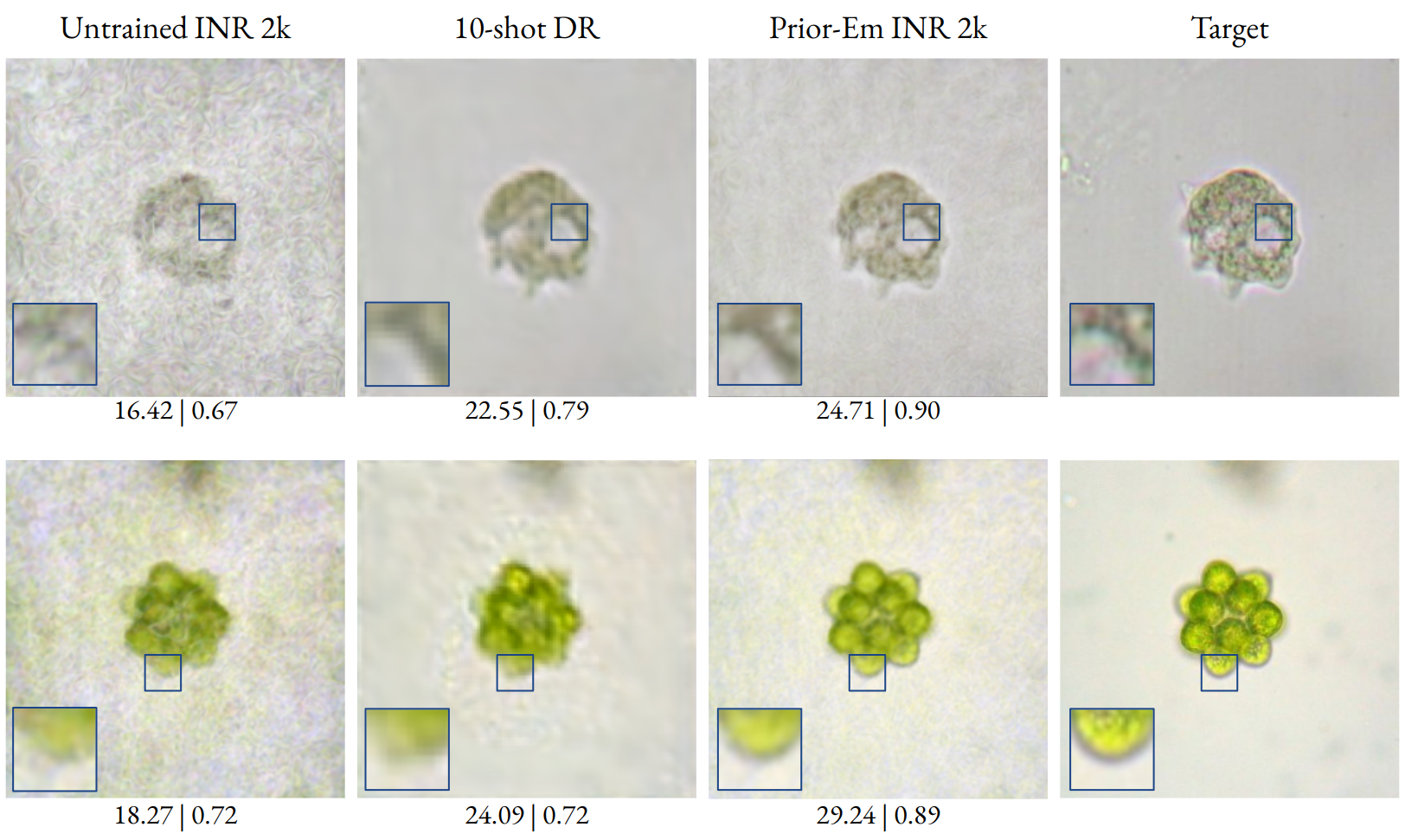}
\caption{Visual comparison of reconstruction results corresponding to prior-embedded INR. The extreme left column displays the untrained INR reconstructions at 2k iterations, followed by the 10-shot domain-restricted reconstruction by \cite{banerjee2023reconstructing}, at 2k iterations. Next are the prior-embedded INR reconstructions, with the target image on the far right. The quantitative scores, presented in PSNR (in dB) | SSIM format, are provided below each image.}
\label{vreslow}
\end{figure}

\subsection{Results for Prior-Embedded Untrained INR}
In this section, we present the results obtained using our method in the prior-embedded scenario. As anticipated, there is a significant improvement in reconstruction performance and convergence speed. For a fair comparison, we selected the approach from \cite{banerjee2023reconstructing}, which performs domain-restricted low-shot reconstruction of lensless images using only around 5 to 10 domain-specific examples. Considering the target area of lensless microscopy, we utilized the microorganism pre-processed images dataset \cite{amoeba} to obtain the low-shot examples.

For the prior-embedded untrained optimization of the INR, we used a single image from the dataset for prior embedding. Then, we selected another image from the same domain and created a synthetic lensless image using the forward model with the known PSF from the DiffuserCam dataset. Subsequently, we performed untrained optimization to obtain a reconstruction with the INR initialized with the prior embedding.

The resulting reconstructions are shown in Fig. \ref{vreslow}. It is evident from the figure that the untrained INR with prior embedding clearly outperforms the 10-shot domain-restricted reconstruction method. This demonstrates the efficacy of incorporating prior knowledge into the INR framework, leading to superior reconstruction quality and faster convergence.

\section{Discussion and Conclusion}
In this work, we explored implicit neural representations (INRs) for lensless image deblurring, specifically comparing untrained INRs against the modified Deep Decoder. We leveraged the concept of prior embedding to enhance reconstruction quality and our findings indicate that INRs offer significant advantages in terms of smoothness, continuity, and the ability to handle sparse and irregularly sampled data.

Our quantitative evaluation, using metrics like PSNR and SSIM, demonstrates that untrained INRs outperform the existing untrained methods by a significant margin in both reconstruction quality and convergence speed. This was evident across multiple parameter counts, where INRs consistently achieved higher fidelity reconstructions with fewer parameters. The concept of under-parameterization was quantified using the Under-Parameterization Ratio (UPR), which allowed us to systematically study the impact of network size on reconstruction performance. Networks with too few parameters struggled to capture structural information, but as the parameter count increased, the reconstructions improved significantly in both structure and color fidelity.

In the case of prior-embedded untrained INRs, we observed a marked improvement in reconstruction quality and faster convergence. By embedding prior information into the network, we were able to guide the optimization process more effectively, leading to more accurate and robust reconstructions. This approach was compared against a domain-restricted low-shot method from \cite{banerjee2023reconstructing}, where prior-embedded INRs showed clear superiority in performance.

Visual comparisons with other established methods, such as ADMM \cite{boyd2011distributed} and Learned-ADMM-UNet \cite{monakhova2019learned}, further validated the effectiveness of our approach. While ADMM performed poorly in comparison, Learned-ADMM-UNet showed slightly better structure restoration in some cases. Nonetheless, untrained INRs consistently provided superior overall reconstruction quality.

The integration of the physics-informed forward model, utilizing the known PSF in the optimization loop, was crucial for achieving high-quality reconstructions. This approach ensures that the network parameters are updated in a manner that respects the underlying physical imaging process, further enhancing the fidelity of the reconstructed images. Our study highlights the potential of untrained INRs for lensless image deblurring, particularly when combined with prior embedding. The ability to achieve high-quality reconstructions with fewer parameters and faster convergence makes INRs a promising avenue for future research in computational imaging. Future work could explore further optimizations in network architecture and embedding strategies to push the boundaries of what is possible with INRs in various imaging applications.

\bibliographystyle{ieeetr}
{\bibliography{bib}}


\end{document}